\documentstyle[12pt,frascatiphys]{article}

\title{Multichannel Interaction Analysis of Scalar Mesons Below 1800 MeV}
\author{R. Kami\'nski$^{1,2}$, L. Le\'sniak$^1$ and B.\ Loiseau$^2$\\ 
{\em $^1$Department of Theoretical Physics}\\
{\em The Henryk Niewodnicza\'nski Institute of Nuclear Physics,}\\ 
{\em PL 31-342 Krak\'ow, Poland}\\
{\em $^2$LPTPE Universit\'e P. et M. Curie, 4, Place Jussieu,} \\
{\em              75252 Paris CEDEX 05, France}}

\newcommand {\be}{\begin{equation}}
\newcommand {\eb}{\end{equation}}
\newcommand{\ba}{\begin{eqnarray}}
\newcommand{\ea}{\end{eqnarray}}
\newcommand{\pp}{$\pi\pi$ }
\newcommand{\roro}{$\sigma\sigma$ }
\newcommand{\kk}{$K\overline{K}$ }
\newcommand{\fo}{$f_0(980)$ }
\newcommand{\epw}{$f_0(1400)$ }
\newcommand{\epsig}{$f_0(500)$ }

\newcommand{\reactpol}{$\pi^- p_{\uparrow} \rightarrow \pi^+ \pi^- n$ }

\begin{document}

 \maketitle

\baselineskip=14.5pt

\begin{abstract}

Properties of scalar--isoscalar mesons in a mass range from 
\pp threshold up to 1800 MeV are analysed using an
unitary
model with separable interactions 
in three decay channels: \pp, \kk and an effective $2\pi2\pi$. 
Different solutions are obtained by fitting \pp and \kk data. 
Ana\-ly\-ti\-cal structure of the
meson-meson multichannel amplitudes is studied with a special emphasis
on the important role played by the $S$-matrix zeroes. 
The dependence of the positions of $S$-matrix singularities  on the
interchannel coupling strength  is investigated.
Poles, located in the complex energy plane not
too far from the physical region, are interpreted as
scalar resonances: a wide \epsig, a narrow \fo and a relatively 
narrow $f_0(1400)$. 
In all our solutions two resonances, lying on different sheets, in the energy 
region between  1300 MeV and 1500 MeV are found.
These states may be compared with the resonances $f_0(1370)$ and $f_0(1500)$ 
seen in the experiments at CERN.
Total, elastic and inelastic channel cross sections, 
branching ratios and coupling constants are evaluated and
compared with available data. 

\end{abstract}

\baselineskip=17pt

\section{Model}

Scalar mesons are still not well enough identified and classified.
Many theoretical and experimental efforts\cite{pdg98,hadron97}
have been recently made for a
better understanding of their dynamics and structure. 
In our analysis we use a separable potential model\cite{kll2}
for three coupled channels: \pp, \kk and \roro (an effective $2\pi2\pi$ channel).   
We solve a system of Lippmann-Schwinger equations with relativistic propagators
in all channels. 
The scattering amplitudes, obtained in such a way, allow us to calculate phase
shifts and inelasticities as functions of 14 potential parameters.
All these parameters are determined from fits to experimental data 
in the \pp and \kk channels.
The \pp data
consist of the 
\pp phase shifts and inelasticities from the CERN-Cracow-Munich measurements 
of the \reactpol reaction on a polarized target\cite{klr}.
We furthermore use the \kk  phase shifts  obtained from reactions on 
unpolarized targets\cite{cohen}.
Before fitting we do not assume anything about possible structures of the
scattering amplitudes.
We find 6 different solutions\cite{kll2,kll} 
with $\chi^2$ values smaller than 116 for 102 degrees of freedom.
Especially good fits with $\chi^2$ smaller than 101 are obtained for
the down-flat set of the \pp data.
We then analyse the analytical structure of the calculated  scattering
amplitudes to find the $S$-matrix poles and zeroes 
in the complex energy plane.

In order to understand the role played by interchannel interactions in the
scalar meson dynamics we first study the analytical structure of the
$S$-matrix elements in the fully decoupled case where all the interchannel
couplings are equal to zero.
In such a case all channels are separated and one can recognise in
which channel particular poles are created.
When the interchannel couplings are switched on, all the poles change positions
and split into poles lying on different sheets.
The sheets are labelled by  $(g_1, g_2, g_3)$ where $g_i$ are
the signs
of the imaginary parts of the complex momenta $k_i$ in particular channels 
($i = 1$ for \pp, $i = 2$ for \kk and $i = 3$ for \roro). 

All diagonal $S$-matrix elements can be expressed as ratios of two
Jost functions $D(k_1,k_2,k_3)$ with minus sign of the momentum $k_i$ in the 
numerator.
Therefore positions of the poles (also in the fully coupled case 
when the interchannel couplings are switched on) are common for all 
the $S$-matrix elements.
However, positions of $S$-matrix zeroes depend on the chosen channel.
Only poles and zeroes which lie close enough to 
the physical region have a significant influence on the phase shifts and 
inelasticities measured experimentally.
We then study poles and zeroes nearby the physical regions and relate them
to the scalar resonances.

\section{Results}

In Table \ref{posa} we show the positions of the $S$-matrix poles for
solution A in the fully decoupled case (column "no couplings") and the fully
coupled case (column "with couplings").
\begin{table}
\centering
\caption{\it Positions of $S$-matrix poles for the solution A (in MeV)} 

\vspace{0.5cm}

\begin{tabular}{|c|c|c|c|c|c|c|}
\hline
channel & \multicolumn{2}{|c|}{no couplings} &
\multicolumn{2}{|c|}{with couplings} & Sign of & \\
\cline{2-5}
& \multicolumn{1}{|c|}{$Re E$} &
  \multicolumn{1}{|c|}{$Im E$} &
  \multicolumn{1}{|c|}{$Re E$} &
  \multicolumn{1}{|c|}{$Im E$} & $Imk_{\pi}$ $Imk_K$ $Imk_{\sigma}$ 
                               & No \\
 \hline
&&      &   564 & $-279$ & $- - -$ & I \\
&&      &   518 & $-261$ & $- + +$ & II \\
$\pi\pi$&   658 & $-607$ &   211 &    0   & $- + -$ & III \\
&&      &   532 & $-315$ & $- - +$ & IV \\
&&      &   235 &    0   & $+ + -$ & V \\
\hline
&&      &  1405 &  $-74$ & $- - -$ & VI \\
$\pi\pi$&  1346 & $-275$ &  1445 & $-116$ & $- + +$ & VII \\
&&      &  1424 &  $-94$ & $- + -$ & VIII \\
&&      &  1456 &  $-47$ & $- - +$ & IX \\
\hline
&&      &  170 &  0 & $+ - -$ & X \\
&&      &  159 &  0 & $- - -$ & XI \\
$K\overline{K}$&  881 & $-498$ &  418 &  $-10$ & $- - +$ & XII \\
&&      & 1038 &  $-204$ & $- + -$ & XIII \\
&&      &  988 &  $-31$ & $- + +$ & XIV \\
\hline
&& & 4741& $-4688$ & $- - -$ & XV \\
$\sigma\sigma$ &  118 & $-2227$ & 3687 & $-2875$& $- + -$ & XVI \\
&& & 3626 & $-3456$ & $+ - -$ & XVII \\
&& & 3533 &  $-579$ & $+ + -$ & XVIII \\
\hline
\end{tabular}
\label{posa}
\end{table}
\begin{figure}[t]
\vspace{9.8cm}
\includegraphics{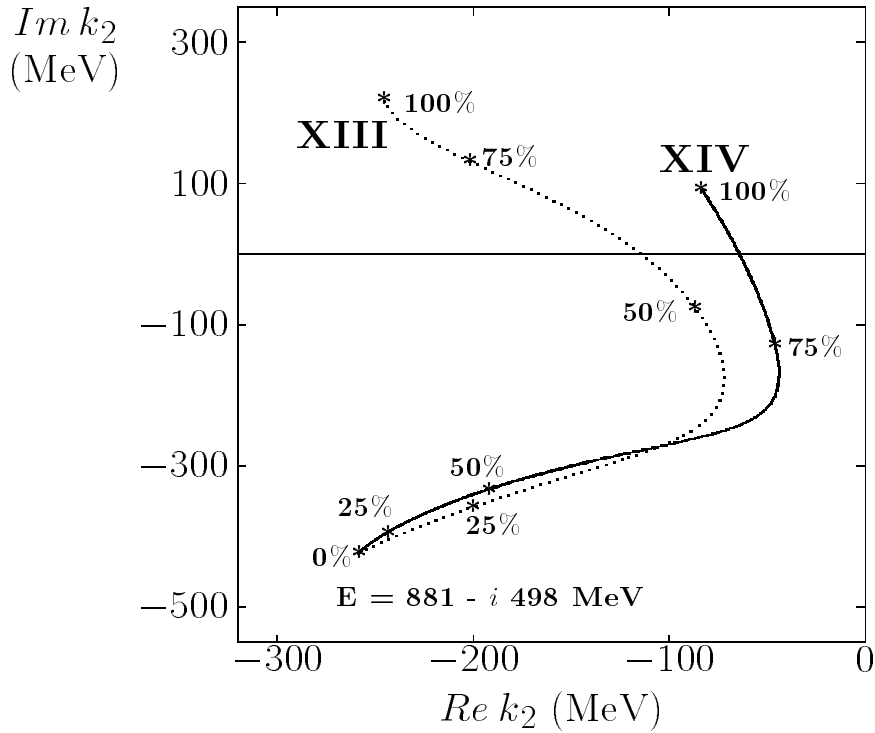}
\caption{\it Pole trajectories in the complex momentum 
plane as a function of the percentage of the interchannel coupling strength 
for the solution A in the \kk channel. 
Roman numbers and energies are taken from Table \ref{posa}
\label{trace}} 
  \end{figure}
\begin{table}
\centering
\caption{\it Average masses and widths of resonances \epsig, \fo and 
\epw found in our solutions A, B, E and F. Here errors represent 
the maximum departure from the average.} 

\vspace{0.3cm}

\begin{tabular}{|c|c|c|c|}
\hline
resonance & mass (MeV) & width (MeV) & sheet \\
\hline 
\epsig or $\sigma$ & $523 \pm 12$ & $518 \pm 14$ & $-++$ \\
\hline
\fo & $991 \pm 3$ & $71 \pm 14$ &  $-++$ \\
\hline
& $1406 \pm 19$ & $160 \pm 12$ & $---$ \\
\epw & $1447  \pm 27$ & $108 \pm 46$ & $--+$ \\
\hline 
\end{tabular}
\label{resonances}
\end{table}
When the interchannel couplings are switched on, starting from the fully
decoupled case, the pole in the \kk channel at $E = 881 -i488$ MeV
splits and moves, among others, to two poles XIII and XIV.
As can be seen in Fig. \ref{trace} both those poles
change sheets when the interchannel couplings are increasing.
Pole XIV on sheet $(-++)$ lies closer to the physical region than 
pole XIII on sheet $(-+-)$ and is related to \fo state.
For the solutions E and F the corresponding poles lie, in the fully
decoupled case, on the positive part of the imaginary axis in the kaon
complex momentum plane and therefore they correspond to \kk bound states.
When the interchannel couplings are switched on then these poles move away from
the imaginary axis creating so called quasi-bound states. 
In solutions A, B, C and D state \fo appears as an ordinary resonance
in absence of the interchannel interactions.
Unambiguous determination of the nature of the \fo state requires
new precise experimental data on \kk interactions near threshold.
    
\begin{figure}
                \vspace{6cm} 
\includegraphics{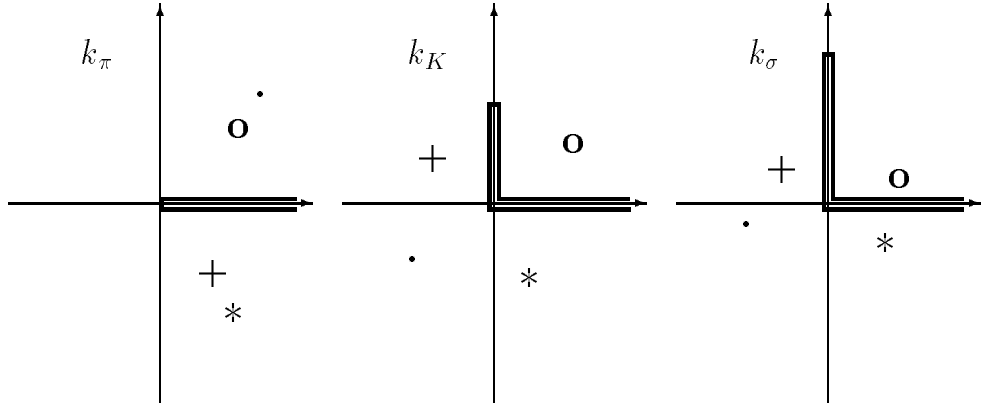}

\caption{\it Planes of complex momenta with schematic positions of the 
$S_{\pi\pi}$ poles (denoted by $*$ on sheet $---$ and by \mbox{\bf +} 
on sheet $-++$) and 
zeroes (denoted by {\bf $^.$} on sheet $+--$ and by {\bf o} on sheet $+++$).
Solid double lines denote physical regions in particular channels.
\label{plane}} 
  \end{figure}

In Table \ref{resonances} we  present masses and widths of three scalar 
resonances found in all our solutions.
Particular choice of the sheets for poles is related to positions of both 
the $S$-matrix poles and zeroes.
An example of such a choice is shown in Fig.\ref{plane} for different
complex momenta planes. 
We there present positions of the poles on sheets $(---)$ and
$(-++)$ together with the corresponding zeroes on sheets $(+--)$ and $(+++)$
respectively.
In the pion complex momentum plane the poles lie below the real 
axis $(Imk_1 < 0)$ close to the physical region.
The sheets for the zeroes are determined for the $S_{\pi\pi}$-matrix
element.
From  the relation $S(k_1,k_2, k_3) = S^*(-k_1^*, -k_2^*, -k_3^*)$ we infer
that all
the poles and zeroes in Fig. \ref{plane} have their twins (not indicated in
figure) which lie symmetrically with
respect to the imaginary axes.
Below 1 GeV the poles on sheet $(-++)$ and the corresponding zeroes on sheet 
$(+++)$ lie close to the physical regions in all channels
and are related to the \epsig and \fo states.

In vicinity of the \roro threshold the two poles on sheets $(---)$ and
$(--+)$, with the corresponding zeroes on sheets $(+--)$ and $(+-+)$ respectively,
lie close to the physical regions in all channels.
We call both of them  \epw. 
Existence of two nearby states in the energy range between 1300 and 1500 MeV
was recently considered in many phenomenological works\cite{pdg98}.
In our analysis an appearance of these states in the fully coupled case is due to the 
splitting of one single pole in the fully decoupled case.

\begin{figure}
\vspace{15cm}
\includegraphics{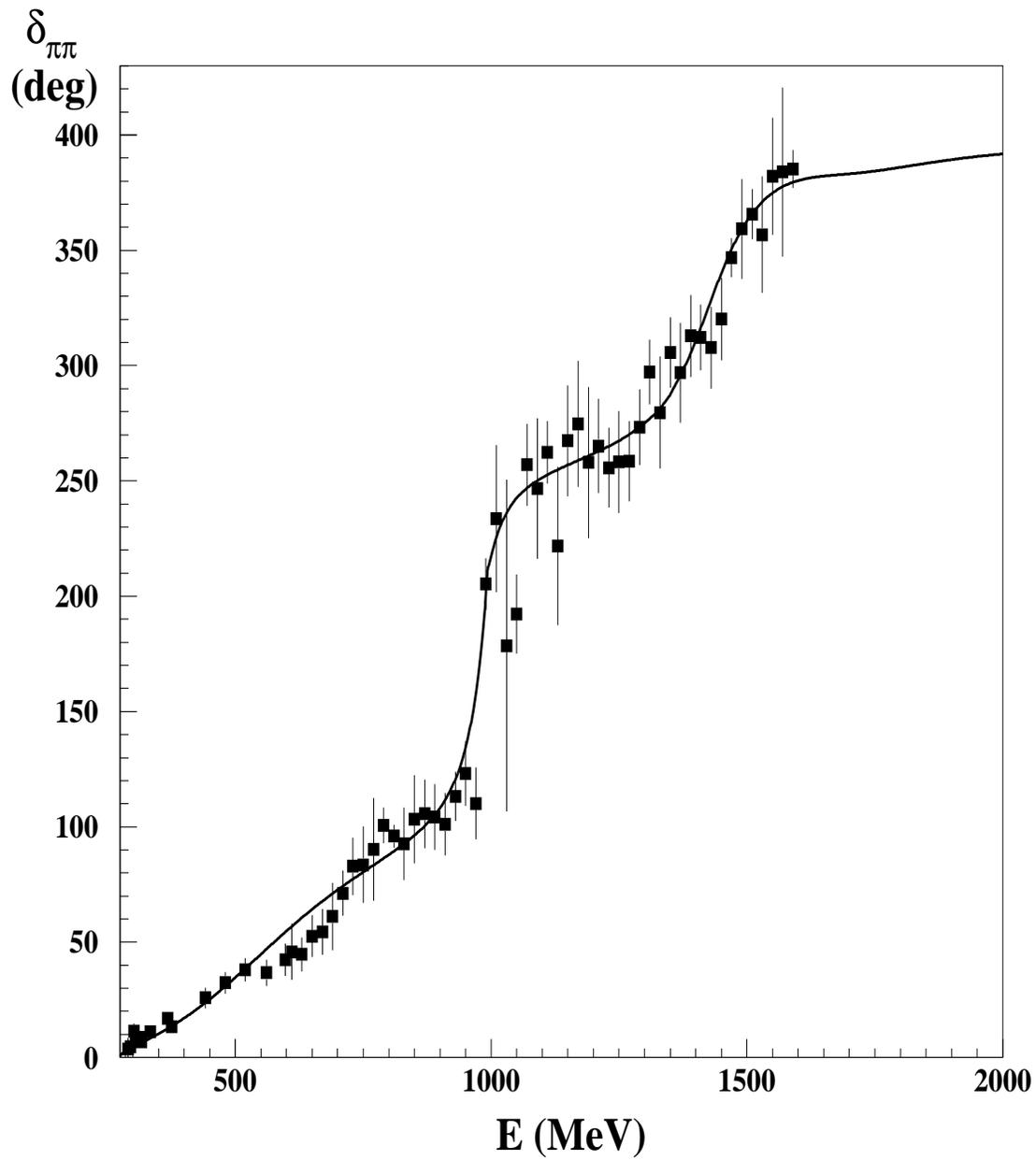}

\caption{\it Energy dependence of \pp phase shifts for the solution E
\label{ephase}} 
  \end{figure}

\begin{figure}
\vspace{10cm}
\includegraphics{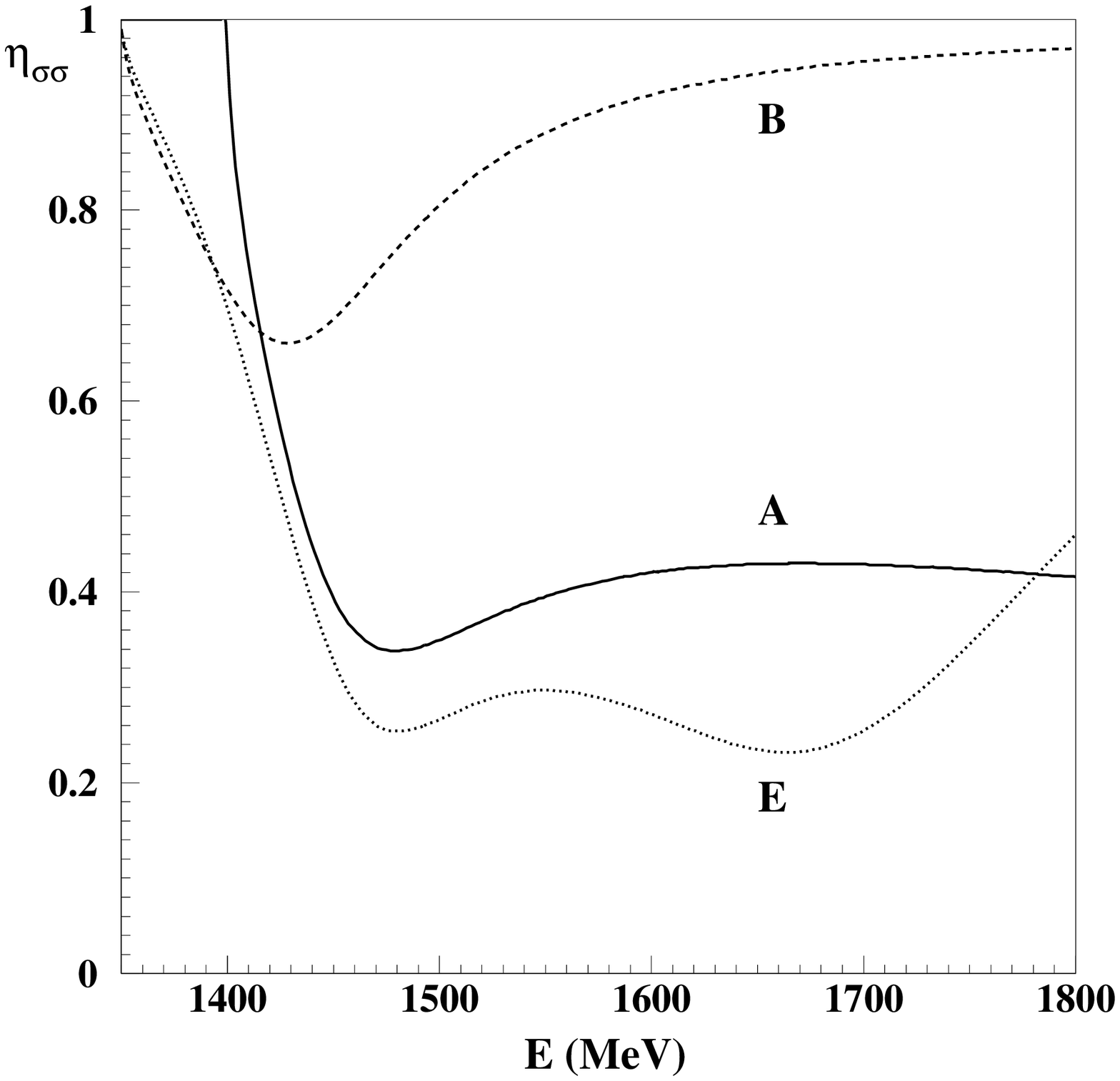}

\caption{\it Energy dependence of the \roro inelasticity  for the
solutions A, B and E \label{eta}} 
  \end{figure}

An interesting  example of the interplay of the $S$-matrix poles and zeroes
can be found in solution E. 
In Table \ref{interplay} we present the set of poles lying close to
the energy region 1600--1700 MeV for this solution.
All poles come from  splitting of the single pole in the \roro channel
at $E = 1565 - i112$ MeV.
As a result of the interchannel interactions  
the pole XIII and the $S_{\pi\pi}$ zero corresponding 
to the pole XVI lie on the same sheet $(---)$.
As it can be seen in Fig. \ref{ephase}, a flat dependence of the \pp phase
shifts above 1600 MeV is the consequence of a strong mutual cancellation of
the nearby pole and the zero.
In Fig. \ref{eta} their influence on the \roro scattering amplitude can be,
however, noticed in the energy dependence of \roro inelasticities.
For the solution E  the two minima correspond to the states \epw
and $f_0(1700)$. In solutions A and B there is only one minimum related to 
\epw.

\begin{table}
\centering
\caption{\it Positions of the $S$-matrix poles around 1600--1700 MeV 
for the solution E (in MeV)} 

\vspace{0.5cm}

\begin{tabular}{|c|c|c|c|c|c|c|}
\hline
channel & \multicolumn{2}{|c|}{no couplings} &
\multicolumn{2}{|c|}{with couplings} & Sign of & \\
\cline{2-5}
& \multicolumn{1}{|c|}{$Re E$} &
  \multicolumn{1}{|c|}{$Im E$} &
  \multicolumn{1}{|c|}{$Re E$} &
  \multicolumn{1}{|c|}{$Im E$} & $Imk_{\pi}$ $Imk_K$ $Imk_{\sigma}$ 
                               & No \\
 \hline
&&       & 1703 & $-271$ & $- - -$ & XIII\\
$\sigma\sigma$ &  1565 & $-112$ & 1648  & $-67$ & $- + -$ & XIV\\
&&       & 1673  &  $-77$ & $- - +$ & XV\\
&&       & 1624 & $-175$ & $+ - -$ & XVI\\
\hline
\end{tabular}
\label{interplay}
\end{table}

In\cite{kll2} together with total, elastic and inelastic
cross sections we present the branching 
ratios calculated for the different channels.
Above the \roro threshold we use $3\times3$ matrix
in which each row describes the three branching ratios in each channel.
We find that the phenomenological information contained in such a matrix
is much reacher than in the Breit-Wigner approach where only three
branching ratios can be used for the three channel case.
We perform our calculations in the energy regions near the \fo and
\epw resonances and also for energies between them.
We also give formulae for the coupling constants and compare their different 
approximations.
We find that the \fo couples strongly to \kk  and the \epw to \pp  channels. 
Coupling of the \epw to the \roro channel can be also significant.

\section{Conclusions}

A proper choice of not only the $S$-matrix poles but also of its zeroes plays 
a crucial
role in the determination of the scalar meson parameters.
It is difficult to find unambiguous values of these parameters
looking only at the energy dependence of phase shifts and inelasticities.
The physical interpretation becomes especially complicated when
the interchannel couplings
are strong and lead to large movements of poles and zeroes in the complex
energy plane.
The positions of both the $S$-matrix poles \underline{and} zeroes have
a significant influence on the experimentally measured observables.
Analysis of the analytical structure of scattering amplitudes is
necessary to determine the nature of scalar mesons.

 


\begin{thebibliography}{99}
\bibitem{pdg98} Particle Data Group, C. Caso {\em et al.},
  Europ. Phys. J. {\bf C3}, 1  (1998).

\bibitem{hadron97} Proceedings of the Hadron Spectroscopy Conference, 
Upton, NY, August 1997, (ed. S.-U. Chung and H. J. Willutzki,
American Institute of Physics, Woodbury, New York 1998). 


\bibitem{kll2} R. Kami\'nski, L. Le\'sniak and B. Loiseau, hep-ph/9810386,
to appear in Europ. Phys. J. {\bf C}.

\bibitem{klr} R. Kami\'nski, L. Le\'sniak, and K. Rybicki, 
              Z. Phys {\bf C74}, 79 (1997).

\bibitem{cohen} D. Cohen {\em et al.}, Phys. Rev. {\bf D22}, 2595 (1980).

\bibitem{kll} R. Kami\'nski, L. Le\'sniak and B. Loiseau, Phys. Lett. {\bf
B413}, 130 (1997).



\end{thebibliography}
\end{document}